\documentclass[11pt]{article}

\usepackage[T1]{fontenc}
\usepackage[utf8]{inputenc}
\usepackage[english]{babel}
\usepackage{mathptmx}% Times text + math; present in every TeX Live, incl. arXiv's
\usepackage[margin=1in]{geometry}

\usepackage{graphicx}
\usepackage{booktabs}
\usepackage{longtable}
\usepackage{multicol,multirow}
\usepackage{amsmath,amssymb,amsfonts}
\usepackage{mathrsfs}
\usepackage{courier}
\usepackage{amsthm}
\usepackage{rotating}
\usepackage{array}
\newcolumntype{L}[1]{>{\raggedright\arraybackslash}p{#1}}% ragged-right wrapped column for text-heavy tables
\newcolumntype{C}[1]{>{\centering\arraybackslash}p{#1}}% centered wrapped column for numeric/short-entry tables
\usepackage[authoryear]{natbib}
\setcitestyle{aysep={}}% no comma between author and year: (Author Year)
\usepackage{textcomp}
\usepackage{xcolor}
\usepackage{caption}
\usepackage[colorlinks=true,allcolors=blue!55!black,breaklinks=true]{hyperref}

\newcommand{\TBL}[1]{#1}                 % table caption wrapper
\newcommand{\TCH}[1]{\textbf{#1}}        % table column head
\newcommand{\botrule}{\bottomrule}

\newcommand{\orcidlink}[1]{%
  \href{https://orcid.org/#1}{\textcolor{orange!85!black}{\textsuperscript{\tiny\textbf{iD}}}}}

\usepackage{titling}
\usepackage{fancyhdr}
\fancypagestyle{plain}{\fancyhf{}\fancyfoot[C]{\small\thepage}}

\begin{document}

%% ===========================================================================
%% FRONTMATTER
%% ===========================================================================
\begin{titlepage}
\thispagestyle{empty}
\vspace*{-1.4cm}
\begin{center}

{\LARGE\bfseries Crowd-Sourced Geographies of Income:\\[0.3em]
Using Google Maps Points of Interest as High-Frequency\\[0.2em]
Proxies for Sub-Municipal Income Estimation\\[0.2em]
in S\~{a}o Paulo, Brazil\par}

\vspace{0.8em}

{\large
Adrienne C. Kinney\textsuperscript{1,2}\thinspace*\quad
Anya Workman\textsuperscript{1}\quad
Ademar Takeo Akabane\textsuperscript{3}\\[0.4em]
Jenna Barac\textsuperscript{1}\quad
Paulo Fernando Braga Carvalho\textsuperscript{4}\quad
Jeova Farias\textsuperscript{5}\\[0.4em]
Fernando Nascimento\textsuperscript{1,2}\quad
Paulo Ricardo da Silva Oliveira\textsuperscript{6}
\par}

\vspace{0.7em}

{\footnotesize\begin{minipage}{0.88\textwidth}\centering
\textsuperscript{1}Hastings Initiative for AI and Humanity, Bowdoin College, Brunswick, Maine, United States\\
\textsuperscript{2}Department of Digital and Computational Studies, Bowdoin College, Brunswick, Maine, United States\\
\textsuperscript{3}Postgraduate Program in Urban Infrastructure Systems, Polytechnique School, Pontifical Catholic University of Campinas, Campinas, S\~{a}o Paulo, Brazil\\
\textsuperscript{4}Postgraduate Program in Geography -- Spatial Information Treatment, Pontifical Catholic University of Minas Gerais, Belo Horizonte, Minas Gerais, Brazil\\
\textsuperscript{5}Department of Computer Science, Bowdoin College, Brunswick, Maine, United States\\
\textsuperscript{6}Center for Applied Research in Economics and Society (Hub Social, Observat\'{o}rio PUC-Campinas), School of Economics and Business, Pontifical Catholic University of Campinas, Campinas, S\~{a}o Paulo, Brazil\\[0.5em]
\textsuperscript{*}Corresponding author: \href{mailto:a.kinney@bowdoin.edu}{a.kinney@bowdoin.edu}
\end{minipage}\par}

\end{center}

\vspace{0.6em}

\begin{center}
\begin{minipage}{0.92\textwidth}
{\centering\textbf{Abstract}\par}
\vspace{0.3em}
\small
Accurate, up-to-date income data at the sub-municipal scale is essential for social policy
in middle-income countries, yet in Brazil it depends on a costly decennial census whose intercensal
gap recently exceeded a decade. We test whether the composition of crowd-sourced Google Maps Points of
Interest (POIs) can serve as a high-frequency, low-cost proxy for household income across the 26,625
census sectors of the municipality of S\~{a}o Paulo. Using a theoretically motivated set of POI categories
retrieved from Google Places, we represent each sector by its POI counts, decompose these high-dimensional,
sparse features with principal component analysis (PCA) and non-negative matrix factorization (NMF), and train a sweep of regression models to predict census-derived
income. Under a data leakage-aware spatial validation design the best model (NMF with gradient boosting)
attains a held-out $R^2$ of 0.65, with performance stable across feature-extraction methods.
Interpretable decompositions reveal which POI types carry the income signal. These results suggest
that commercial, crowd-sourced geospatial data can complement conventional income statistics during
intercensal periods, and we discuss extensions toward multidimensional poverty and the capabilities
framework.
\end{minipage}
\end{center}

\vspace{0.5em}

\begin{center}
\begin{minipage}{0.92\textwidth}
{\centering\textbf{Policy Significance Statement}\par}
\vspace{0.3em}
\footnotesize
In many cases governments allocate social programmes using income maps that can be more than a decade old.
Before 2022, S\~{a}o Paulo's most recent census income data dated to 2010. This study shows that the mix of
businesses and amenities listed on Google Maps can predict neighbourhood income across S\~{a}o Paulo's
census sectors, explaining about two-thirds of income variation in previously unseen areas. These data are
inexpensive to obtain and continuously updated, making them a practical option for refreshing income estimates
between censuses, flagging neighbourhoods whose conditions are changing, and cross-checking the administrative registries
used for programme targeting. The method relies on interpretable models and low computational cost, making it
readily adoptable by under-resourced policy teams.
\end{minipage}
\end{center}

\vspace{0.5em}

\begin{center}
\begin{minipage}{0.92\textwidth}
\end{minipage}
\end{center}

\end{titlepage}

\setcounter{page}{1}

\section{Introduction}

Accurate, current sub-municipal income data is a valuable input to social policy in middle-income 
countries, yet traditional census mechanisms rarely provide it at the frequency or timeliness policy 
requires. In Brazil, the primary source of small-area income estimates is the Instituto Brasileiro de 
Geografia e Estat\'{\i}stica (IBGE) Censo Demogr\'{a}fico, which provides a nationally representative enumeration 
conducted, under normal circumstances, every ten years. The most recent complete census was conducted 
in 2010. The subsequent census, delayed first by budgetary constraints and then suspended by the 
COVID-19 pandemic, began fieldwork only in 2022 and has released income data incrementally since 2023. 
The practical consequence for social policy in the municipality of S\~{a}o Paulo (SP) --- with a population
of 11.45 million \citep{ibge2022censo}, and a wide income inequality gradient --- is that programme targeting, 
vulnerability index construction, and social assistance planning have relied on income maps that are, 
at the time of writing, between twelve and fifteen years out of date.

Such a temporal gap has significant consequences for public policy planning. In Brazil, social 
assistance is organised through the Sistema \'{U}nico de Assist\^{e}ncia Social (SUAS), whose units include the 
Centro de Refer\^{e}ncia de Assist\^{e}ncia Social (CRAS), which delivers Basic Social Protection, and the 
Centro de Refer\^{e}ncia Especializado de Assist\^{e}ncia Social (CREAS), which delivers Special Social 
Protection \citep{brasil2009orientacoes,brasil2012norma}. Planning for these units draws on two kinds of data: the Cadastro 
\'{U}nico (Cad\'{U}nico), a continuously updated administrative registry of low-income families, and periodic 
IBGE statistics, primarily the decennial Censo Demogr\'{a}fico and the continuous Pesquisa Nacional por 
Amostra de Domic\'{\i}lios. The two have complementary limitations: the population-representative statistics 
that anchor sub-municipal income estimates age between census waves, while Cad\'{U}nico, though continuously 
updated, reflects only the families that identification efforts have actually reached. Because that 
coverage depends on active outreach, the families in greatest need can remain unregistered where 
outreach is weak, so a continuously updated, full-coverage income proxy is valuable less as a 
replacement for Cad\'{U}nico than as an independent signal of where registration gaps are likely to 
concentrate.

To address this problem, many researchers have been seeking alternatives to provide regularly updated
baseline income information or estimates, part of a broader turn toward data science and machine learning
as inputs across the stages of the policy cycle \citep{anggunia2025decoding}. In that context, geolocated points of interest (POIs), databases 
of named places (businesses, services, institutions) with geographic locations, maintained by platforms 
such as Google Maps, provide a continuously updated, spatially granular alternative signal correlated with 
economic activity and income levels. While a census records household income directly, POI data records the 
institutional and service landscape: the density and composition of banks, employers, transport nodes, shops, 
and services within a territory. 

We develop and test the use of Google Maps POIs for estimating income in the municipality of S\~{a}o Paulo. With
over twenty-six thousand census sectors and wide income variation according to the 2022 census, S\~{a}o Paulo is
a demanding and informative test case for sub-municipal income estimation. We demonstrate that a machine 
learning model trained on these features can predict census-derived income at the sub-municipal scale with 
coefficient of determination ($R^2$) values in the range of 0.62--0.65, providing a low-cost complement to
conventional income data during intercensal periods. 

The next section reviews the growing literature on geolocated and crowd-sourced proxies for income and poverty, 
situating this study within it and identifying the methodological gaps it addresses. We then describe the 
data and the modelling approach, present the results, and discuss the implications of the current study, 
followed by its limitations and future work, including mapping a multidimensional extension, before 
concluding.

\subsection{Geospatial Proxies for Income and Poverty: Literature Review and Methodological Assessment}

Reliable, timely income data remains scarce. According to a World Bank estimate, only 62 out of the 155 
countries studied have poverty data of sufficient frequency to track progress toward development goals 
\citep[p.~21]{serajuddin2015data}. Even in countries with established data collection mechanisms, traditional 
instruments such as household surveys and decennial censuses are costly to administer, slow to produce, and 
difficult to update between enumeration cycles, thus limiting their usefulness. 

The Brazilian IBGE census exemplifies these constraints: at R\$ 2.3 billion ($\approx$ US\$445 million)\footnote[1]{Income
and welfare figures are reported in 2022 international dollars (I\$), using the World Bank Purchasing Power Parity
(PPP) conversion factor for private consumption (R\$2.48 = I\$1); expenditure figures, such as the census cost, are converted
at the 2022 average nominal exchange rate (R\$5.17 = US\$1).} for the 2022 census \citep{ibge2022censo}, more frequent data collection
is financially unfeasible. The postponement of the 2020 census until 2022, resulting in a twelve-year data collection 
gap, further demonstrates the fragility and temporal limitation of these systems. Additionally, coverage and data 
quality vary spatially, with gross enumeration error rates ranging from 8.4\% in Para\'{\i}ba to 22.8\% in Rio de Janeiro \citep[p.~98]{ibge2024pesquisa}.

Taken together, the low frequency rate of data collection, high monetary cost, and spatial coverage unevenness 
limit the census's value as a reliable, timely, or neutral data source for continuous poverty monitoring. Proxy 
measures such as satellite imagery, commercial geospatial databases, and mobile phone records may provide 
cost-effective, timely alternatives to census data for measuring economic conditions.

\subsubsection{The Landscape of Geospatial Income and Poverty Proxies}

Table~\ref{tab1} summarises recent developments in machine learning analysis of geospatial data for income and poverty estimation. 
Together, they suggest a broadly consistent conclusion: geospatial data contains meaningful socioeconomic signals that 
can be leveraged to estimate welfare with reasonable accuracy. Across a range of contexts and validation datasets, 
machine learning models consistently demonstrate promising predictive performance, suggesting that multiple forms of 
data can serve as viable proxies for conventional income and poverty measures. We discuss the appeal and challenges of the 
geospatial data sources in the sections below. 

\begin{table}[!htbp]
\TBL{\caption{Overview of geospatial proxy approaches for income and poverty estimation.\label{tab1}}}
{\small\setlength{\tabcolsep}{3pt}\renewcommand{\arraystretch}{1.25}%
\begin{center}
{%
\begin{tabular}{@{}L{2.0cm}L{2.0cm}L{2.0cm}L{2.2cm}L{3.0cm}L{3.1cm}@{}}\toprule
\TCH{Study} & \TCH{Data Source} & \TCH{Method} & \TCH{Spatial Scale} & \TCH{Validation} & \TCH{Key Limitation} \\\midrule
\citet{jean2016combining} & Satellite imagery (day/night) & CNN + ridge regression & Cluster, approximately villages or wards (Africa) & $R^2$=0.37--0.55 vs LSMS consumption; $R^2$=0.55--0.75 vs DHS asset wealth & Cannot evaluate within-cluster predictions; no temporal validation \\
\citet{blumenstock2015predicting} & Mobile phone metadata (call records) & Ridge regression / elastic net & District and cluster (Rwanda) & r=0.68 vs phone survey, individual wealth; r=0.92 vs DHS district wealth & Requires telco partnership; privacy \\
\citet{chi2022microestimates} & Satellite + mobile + Facebook connectivity & Gradient boosting & LMIC pixel grid & $R^2$=0.56--0.70 vs DHS wealth index; $R^2$=0.72 vs independent census data from LMIC & Composite, hard to attribute contribution \\
\citet{gebru2017using} & Google Street View images & CNN + ridge regression & US ZIP code and precinct & r=0.82 vs ACS household income & Requires large-scale street-level imagery; privacy \\
\citet{steele2017mapping} & Mobile data + satellite & Bayesian geostatistical models & Sub-district (Voronoi polygons) (Bangladesh) & $R^2$=0.76 vs DHS wealth index & Requires telco partnership; performance varies by poverty measure and urban/rural \\
\citet{pokhriyal2017combining} & Mobile CDR + satellite & Gaussian Process regression & Commune (Senegal) & r=0.91 vs census & Requires telco partnership; CDR coverage excludes some; results biased towards urban areas \\
\citet{tingzon2019mapping} & Satellite + OSM POIs & CNN + ridge regression + Random Forest regression & Cluster (Philippines) & $R^2$=0.63 vs DHS wealth index & Accuracy and completeness of OSM coverage \\
This study & Google Places POIs (commercial) & Tree-based regressors & Municipality census sectors (SP-BR) & $R^2$=0.62--0.65 validated vs IBGE income & Formal economy bias; informal gap \\\botrule
\end{tabular}}
\end{center}}
\end{table}

\paragraph{Satellite Imagery}

Historically, night-time lights (NTL) have served as the primary proxy for economic activity, showing strong correlations 
with Gross Domestic Product (GDP) and regional domestic product \citep{perezsindin2021are,castro2022predicting}.
However, NTL data often suffer from coarse resolution, and frequently fail to differentiate economic activity in densely 
populated poor areas and in rural regions with minimal electrification \citep{piaggesi2019predicting,lee2022high}.

Recent work applies machine learning to high-resolution daytime satellite imagery to identify visual features such as 
building types, roof materials, and road networks, often via transfer learning from NTL or ImageNet features 
\citep{tingzon2019mapping,zheng2024county}. Combining NTL with daytime geospatial features, such as land cover, points of 
interest (POIs), and building density, consistently enhances predictive power over any single data source 
\citep{ledesma2020interpretable,zheng2024county,tingzon2019mapping,keola2015monitoring}, and has been used to estimate variables 
including household income, wealth indices, educational attainment, and access to basic utilities \citep{tingzon2019mapping,ledesma2020interpretable}. Feasibility has been demonstrated across diverse geographic contexts, spanning Latin America, Asia, 
and Sub-Saharan Africa \citep{puttanapong2022predicting,perezsindin2021are,castro2022predicting,hu2022village,engstrom2017poverty,lee2022high}. While accuracy is generally higher in urban environments than in rural 
areas, high-resolution poverty mapping now performs reasonably even at village and street-corner levels \citep{hu2022village,lee2022high}.

\paragraph{Mobile Phone and Social Media Data}

The integration of mobile phone data into socioeconomic analysis has provided a cost-effective, granular alternative to 
traditional census surveys \citep{hall2023review,hazem2025multi}. Seminal research in this field established that anonymised 
call detail records can reconstruct high-resolution maps of wealth distribution by analysing historical communication 
patterns and mobility \citep{blumenstock2015predicting,tingzon2019mapping}. Key indicators extracted from mobile metadata 
include call frequency, text messaging history, airtime top-up patterns, and specific handset types, all of which have 
proven accurate predictors of community- and individual-level wealth \citep{blumenstock2015predicting,hall2023review}.

The proliferation of online data from social media, e-commerce, and search platforms offers another perspective for inferring 
socioeconomic variables \citep{dong2019predicting,hall2023review,lamichhane2025exploring}. \citet{ledesma2020interpretable} use Facebook 
advertising data for interpretable poverty mapping, and Google Trends has been used to ``predict the present'' of economic 
status \citep{austin2021using}. \citet{li2019uncovering} show that Foursquare check-in activity across different venue categories such 
as food, nightlife, and travel correlates strongly with median household income. \citet{hazem2025multi} and \citet{ucar2021news} 
analyse mobile application traffic, finding that high-income areas favour information-seeking traffic while lower-income 
regions prefer social media and streaming.

\paragraph{Points of Interest and OpenStreetMap Data}

The use of geolocated POI data for income and poverty estimation is a smaller but growing body of literature. \citet{munetonsanta2023predicting} utilised OSM POIs and land use cover to estimate poverty levels in Medell\'{\i}n, Colombia. \citet{tingzon2019mapping} combined OSM POIs and nighttime lights satellite imagery to map poverty in the Philippines. \citet{lee2022high} used OSM POIs, nighttime luminosity, and population density to estimate village-level poverty across 25 countries 
in Sub-Saharan Africa. These studies share a common theoretical mechanism: POI density and composition proxy for the mix 
of institutions, organizations, and services present in a territory, which in turn correlate with its residents' poverty levels.

Google Places data, the source used in this paper, has been less studied than OSM in the poverty estimation literature, 
largely because OSM's open license reduces access barriers. However, Google Places may be more complete and more consistently 
maintained in urban Brazil than OSM, which has an estimated building-footprint completeness rate of 20\% in Latin American 
and Caribbean urban centres \citep{herfort2023spatio}. Our study is among the first to assess Google Places POI composition 
as a standalone income predictor for Brazilian municipalities and to systematically evaluate which POI categories carry 
the strongest income signal. 

\subsection{Positioning This Study}

The specific contributions of this paper relative to the literature reviewed above are: (1) systematic evaluation of 
Google Places (rather than OSM) POI composition as an income predictor at sub-municipal scale in Brazil; (2) a 
theoretically motivated POI query design, in which categories are chosen for hypothesised income mechanisms rather than 
ingesting all available categories without justification, paired with an interpretable, decomposition-based account 
of which POI types carry the income signal; (3) a rigorous, data-leakage-aware spatial validation design: 
buffered held-out sectors, income-representative train/validation/test splits, bootstrapped confidence intervals on 
held-out performance, and Moran's I diagnostics, which, combined, serve to guard against optimistic performance estimates.

We use Google Places Application Programming Interface (API) as our data source. In Brazil, Google Maps has 
demonstrated superior performance relative to alternative sources. A comparison of geocoding methods applied to a 
Brazilian health surveillance dataset found that Google Maps API had the highest accuracy and coverage, outperforming 
OSM, ArcGIS, and the National Address Database for Statistical Purposes (CNEFE) \citep{sanglard2025use}. The analytical 
value of Google Places POI data is, paradoxically, a by-product of its commercial function. Google Maps is a primary 
digital storefront for businesses operating in Brazil, giving owners a direct incentive to keep listings accurate 
and current. The result is a continuous self-updating record of formal and semi-formal economic activity whose quality 
is, in many urban and peri-urban areas, driven by the intrinsic commercial motivation of business owners and feedback 
from an enormous user base, rather than by a satellite operator's revisit schedule or by volunteer contributors' 
discretion \citep{deri2025crowdsourced}. 

The practical implication of this continuous update mechanism extends beyond measurement. A census-derived poverty 
map is a snapshot that ages silently until the next enumeration. A POI-based proxy, by contrast, is a living signal 
that can be re-queried monthly, quarterly, or in response to a specific economic shock, yielding a revised income-deprivation 
map within days at negligible cost. It also enables impact assessment: with sequential POI snapshots, 
an evaluator can track the predicted income distribution before and after an intervention, providing a low-cost, 
spatially granular complement to survey-based difference-in-differences designs.

Continuous monitoring is of limited use to policymakers if outputs are not interpretable. Although deep learning 
models may outperform simpler models in benchmark comparisons, they are difficult to interpret and validate due to 
their ``black box'' nature \citep{puttanapong2022predicting,ledesma2020interpretable}. The need for explainable models is
particularly acute in public-policy settings, where interpretability supports human oversight and accountable, high-stakes
decisions \citep{amarasinghe2023explainable}. Simpler models offer a critical advantage:
feature importance scores provide a readily interpretable account of which POI types drive predictions, letting
policymakers assess whether the model's reasoning is plausible and allowing researchers to connect findings to economic
theory. We prioritise interpretability by using tree-based models and decomposed, low-dimensional input features. This 
approach also requires fewer computational resources compared to deep learning or image-based approaches, making POIs more 
scalable and accessible to under-resourced policy teams.

\section{Data and Methods}\label{sec:methods}

\subsection{Income from IBGE Census}

\begin{figure}[htbp!]
\centering
\includegraphics[width=\linewidth]{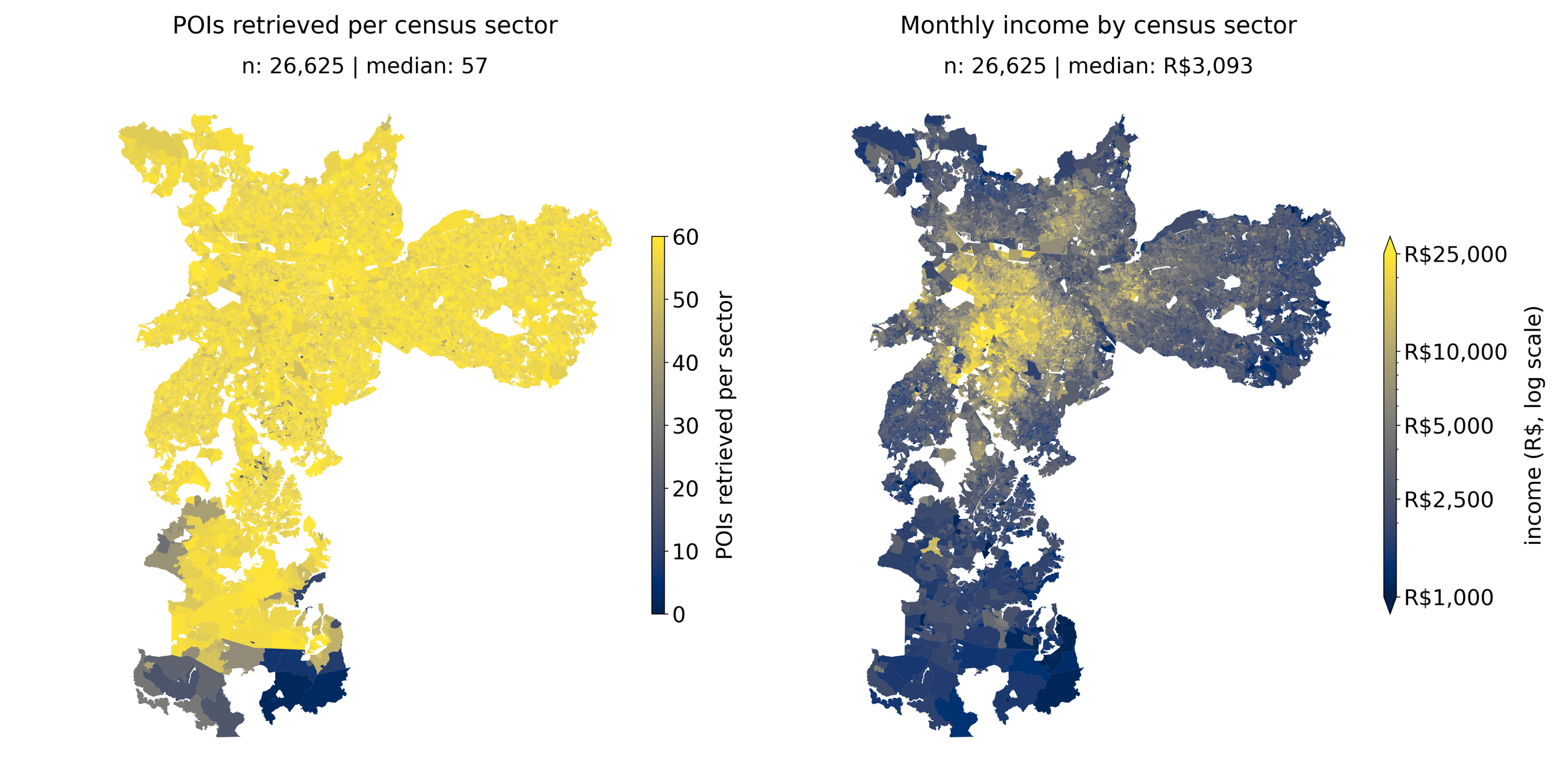}
\caption{(Left) POIs retrieved per census sector across S\~{a}o Paulo municipality (2022 IBGE geometry; $n=26{,}625$ sectors). The colour scale is capped at the per-sector target of 60 POIs. (Right) Monthly household reference person income at the census-sector level (median $\approx$\,I\$1,250/month in 2022 PPP). Colour scale is plotted on a log scale and clipped to the 1st and 99th percentiles of the city distribution\label{fig1}}
\end{figure}

The 2022 IBGE Census provides average gross nominal monthly income received by the household reference person for each 
census sector; affluent sectors concentrate in the urban centre (Figure~\ref{fig1}, right). The POI data were collected in 
April and May of 2026, after the 2022 census reference date.

\begin{figure}[htbp!]
\centering
\includegraphics[width=\linewidth]{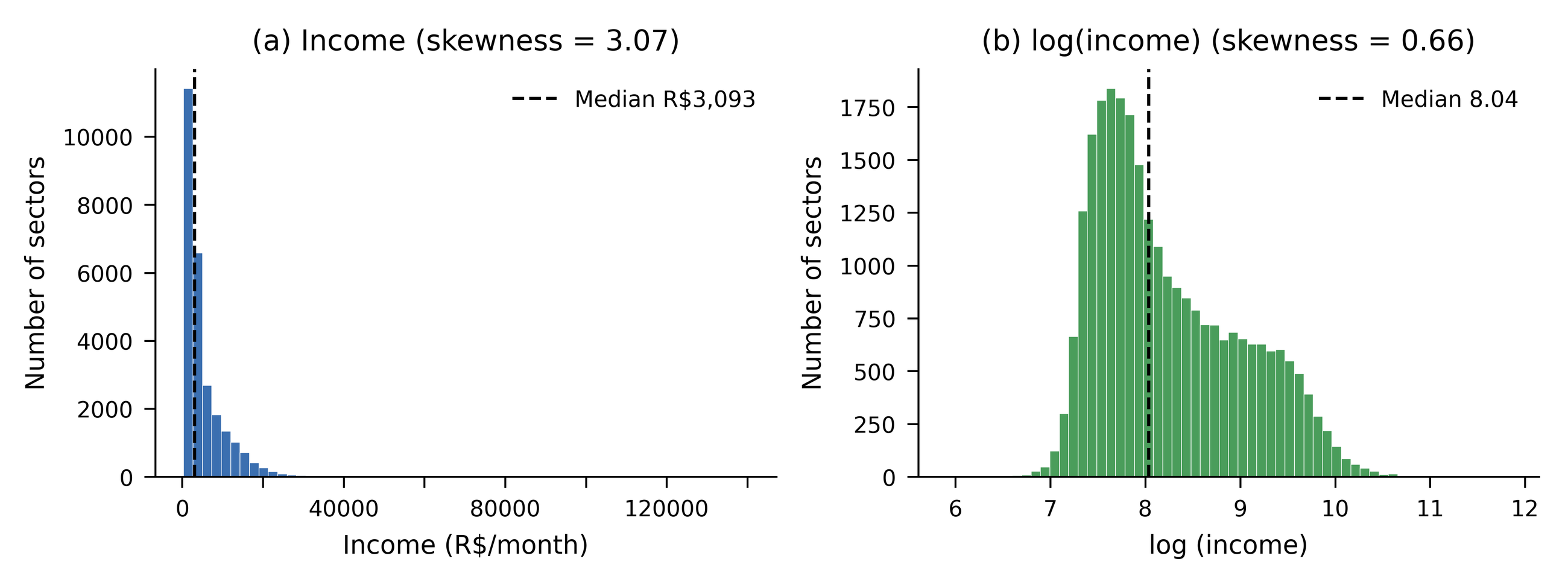}
\caption{Distribution of income across the census sectors of S\~{a}o Paulo. (a) Raw monthly income in Brazilian reais (range R\$367--R\$140,173, median R\$3,093; $\approx$\,I\$148--I\$56,500, median $\approx$\,I\$1,250 in 2022 PPP) and (b) Natural log of income\label{fig2}}
\end{figure}

The household reference person income is strongly right-skewed across S\~{a}o Paulo municipality's 26,625 sectors, ranging
from R\$367/month ($\approx$\,I\$148) to R\$140,173/month ($\approx$\,I\$56,500) with a median of R\$3,093/\allowbreak month ($\approx$\,I\$1,250)
(Figure~\ref{fig2}a, skewness $\approx$\,3.1). On this raw scale, a linear regression model would be dominated by a small number of 
high-income sectors, treating a R\$1,000 error at R\$2,000/month (50\% relative) as equivalent to a R\$1,000 error 
at R\$50,000/month (2\% relative). Taking the natural log compresses the tail, brings the skewness down to 0.66, 
and yields an approximately bell-shaped distribution (Figure~\ref{fig2}b) on which regression residuals are interpretable 
as proportional errors \citep{box1964analysis}. We therefore fit all downstream models to the log of the household 
reference person income and report back-transformed error metrics in raw R\$ as a secondary metric. Hereafter, 
we refer to household reference person income as ``income'' and to its natural log as ``log income.''

\newpage

\subsection{Points of Interest from Google Places}

We employed the Google Places API to retrieve POI data for the municipality of S\~{a}o Paulo (IBGE code 3550308), which
comprises 26,625 census sectors per the 2022 IBGE geometry. Each POI is labelled with a primary type. To capture 
broad commercial diversity, we selected 47 ``anchor'' primary types, based on hypothesised relevance to predicting 
income (Table~\ref{tab2}), partitioned into three groups solely to distribute a set per-sector query budget evenly across the 
taxonomy. For each census sector, we queried up to 20 POIs from each group within a 2 km radius of the sector centroid, 
capped at 60 POIs per sector. Per-sector counts range from a minimum of 2 to the 60-POI cap, with a median of 57, and 
the sectors with lower POI counts are concentrated in the peripheral southern part of the municipality (Figure~\ref{fig1}, left). 
Although we queried only the anchor types, Google's finer taxonomy returned 386 distinct subtypes, so the anchors were 
sufficient to surface a much broader feature space.

{\small\renewcommand{\arraystretch}{1.25}%
\setlength\LTcapwidth{\textwidth}%
\setlength\LTcapwidth{\textwidth}%
\begin{longtable}{@{}L{5.2cm}L{3.8cm}L{5.4cm}@{}}
\caption{Google Places POI categories used as income predictors: selection rationale, expected income mechanism, and data quality caveats.\label{tab2}}\\
\toprule
\TCH{POI Category (Google Places types)} & \TCH{Hypothesised Income Mechanism} & \TCH{Data Quality Note / SP-Specific Caveat} \\
\midrule
\endfirsthead
\caption[]{Google Places POI categories used as income predictors (continued)}\\
\toprule
\TCH{POI Category (Google Places types)} & \TCH{Hypothesised Income Mechanism} & \TCH{Data Quality Note / SP-Specific Caveat} \\
\midrule
\endhead
\midrule
\multicolumn{3}{r}{\small\itshape Continued on next page}\\
\endfoot
\bottomrule
\endlastfoot
Banks, ATMs, credit unions (bank, atm) & Financial inclusion; formal capital access & May be sparse in peripheral S\~{a}o Paulo; digital banking services may not be represented \\
Retail and grocery (supermarket, grocery\_store, store, shopping\_mall) & Consumer spending power; formal retail access & Informal markets (feiras, street vendors, and sacoleiros) may not be captured \\
Education (preschool, primary\_school, secondary\_school, university, library) & Labour market access; human capital development & Education POIs are generally expected to be complete \\
Public transport (train\_station, bus\_station, taxi\_stand, airport, international\_airport) & Commuting capacity; labour market reach & Public transportation expected to be well-covered; informal transportation services may be absent \\
Restaurants and caf\'{e}s (restaurant, cafe, coffee\_shop, bar, pub, sandwich\_shop) & Disposable income signal; urbanisation proxy & Tourism areas may inflate POI density \\
Personal services (beauty\_salon, barber\_shop, hair\_care, nail\_salon) & Consumer expenditure capacity; discretionary spending; context dependent income signal & Small independent businesses may be underrepresented outside city-centre \\
Recreation and culture (art\_gallery, playground, stadium, swimming\_pool, athletic\_field) & Disposable income; quality of life; municipal investment & Public recreational facilities may be well mapped; private clubs and facilities may be incomplete \\
Healthcare (hospital, doctor, dentist, pharmacy, drugstore) & Private spending capacity; healthcare accessibility & Distinguish public Sistema \'{U}nico de Sa\'{u}de from private convenio where possible, as they reflect different income signals \\
Hotels and accommodation (hotel, inn, lodging) & Tourism / business activity; local economic vitality & \\
Transportation services (gas\_station, parking) & Motorisation proxy & \\
Places of worship (church) & Congregation density; community cohesion; income signal context dependent & Does not distinguish between large establishments with small congregations \\
Emergency and civic services (police, fire\_station, post\_office, government\_offices) & Public service provision; deprivation indicator (context dependent) & Reflects government service distribution \\
Agriculture (farm) & Rural economic activity; agricultural employment & Coverage may be inconsistent \\
\end{longtable}}

We then created a POI count vector for each sector by counting the number of POIs per Google type, yielding a 26,625 $\times$ 
386 sparse count matrix. This design has two important properties. First, input samples represent the mix of POIs within 
the 2 km neighbourhood around the sector centroid, not the density of POIs in the sector. Second, because we queried by 
radius around centroids, the windows of neighbouring sectors overlap, particularly in the urban areas with small-area sectors. 
The POI overlap is a key motivation for the spatial validation design.

\subsection{Modelling Approach}

\subsubsection{Spatial Validation Design}

Motivated by the spatial structure of the data, we implemented a spatial validation design that minimises both the income 
bias across the training, validation, and held-out test sectors and the data leakage from the neighbourhood querying process. 
We withheld two stripes containing approximately 15\% of the sectors (n=3,550), randomly choosing the location and rotation 
of the stripes to ensure the income distribution of the held-out sectors is representative of the city-wide distribution, 
as measured by the maximum absolute per-quantile income bias:

\begin{equation}
    \text{max pct bias} = \max_{q \in Q} \left| \frac{F^{-1}_{\text{held-out}}(q) - F^{-1}_{\text{city}}(q)}{F^{-1}_{\text{city}}(q)} \right|
\end{equation}

where $F^{-1}$ is the quantile function and $Q = \{5, 10, 25, 50, 75, 90, 95, 99\}$. We re-rolled the held-out 
stripe locations until the max-quantile bias fell close to 5\%.

\begin{figure}[htbp]
\centering
\includegraphics[scale=0.12]{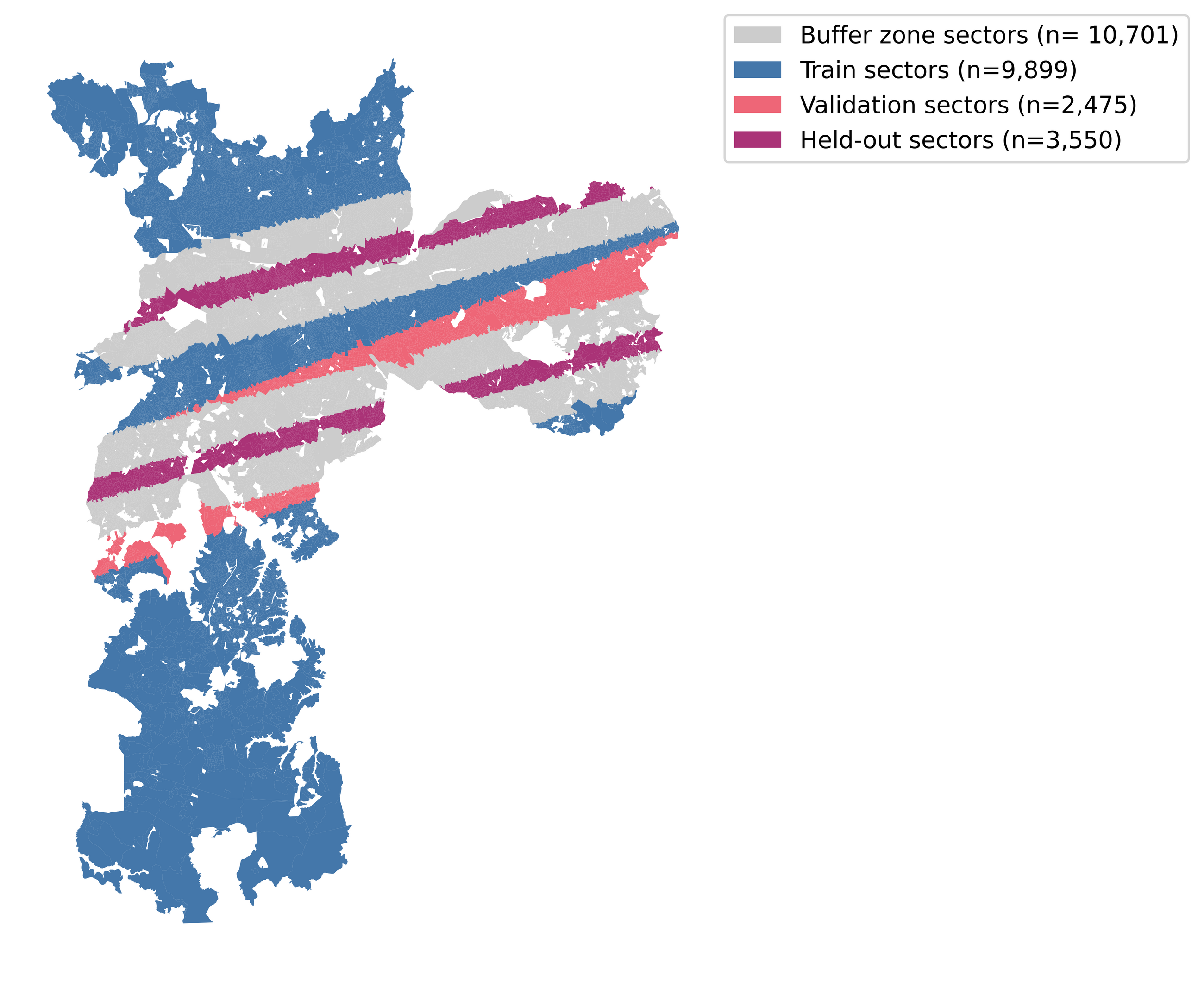}
\caption{Modelling roles across sectors for the spatial validation design. The held-out test set (magenta) is two parallel stripes at a randomly chosen orientation; the buffer zone (grey) is the non-held-out sectors within 3 km of the test stripes, excluded from both training and testing. The remaining pool is split into a training set (blue) and a validation set (pink)\label{fig3}}
\end{figure}

To avoid data leakage from the training set, we buffered the held-out sectors by 3 km and excluded any sectors within 
the buffer from both the training and test sets. From the remaining pool, we selected a validation stripe containing 20\% 
of the training sectors, selecting the stripe location and rotation to minimise income bias versus the city distribution. 
Figure~\ref{fig3} shows the resulting split and Table~\ref{tab3} provides the bias and overlap for all three sets. The 3 km buffer minimised 
overlap without removing as many sectors as a larger buffer width would, and the location and rotation of the validation 
stripe was the best achievable for a single contiguous stripe across 100 re-rolls.

\begin{table}[!htbp]
\begin{center}
\TBL{\caption{Income bias and POI overlap for the training, validation, and held-out test sets.\label{tab3}}}
{\small\renewcommand{\arraystretch}{1.25}%
{%
\begin{tabular}{@{}L{2.0cm}C{1.8cm}C{3cm}C{2.9cm}C{2.9cm}@{}}\toprule
\TCH{Set} & \TCH{Sectors (n)} & \TCH{Income bias vs. city (\%)} & \TCH{POI overlap with training set (\%)} & \TCH{POI overlap with held-out set (\%)} \\\midrule
Train & 9,899 & 14.4 & -- & 1.4 \\
Validation & 2,475 & 18.7 & 67.0 & 3.5 \\
Test (held-out) & 3,550 & 5.2 & 3.7 & -- \\\botrule
\end{tabular}}}        
\end{center}
\end{table}

\subsubsection{Feature Decomposition}

The POI count input matrix is high-dimensional and sparse, with 386 unique POIs and a median of 57 POI counts per sector. 
To mitigate the risk of overfitting and to explore the interpretability of the feature space, we applied two decomposition methods 
to the sectors: principal component analysis \citep{hotelling1933analysis} and non-negative matrix factorisation \citep{lee1999learning}.

Both methods fit a low-dimensional model to the raw features but encode different assumptions about the structure of POI 
co-occurrence. Each finds latent factors that capture patterns of occurrence across the raw POI features. A sector's embedding 
in the component space is a vector of scores that indicate how much of each component is present in the sector, and each component's 
loadings are the weights of the raw POI types that define it. The scores and loadings are interpreted differently across the 
methods, depending on the constraints each method imposes. We fit each decomposition on the training pool, and applied it to 
the held-out test sectors during the prediction stage.

PCA treats the input vectors as a 386-dimensional data cloud and finds orthogonal axes ordered by the variance they explain, 
so the first component captures the largest variance in the data. Since PCA maximises variance on the raw feature scale, a 
few high-count types would dominate the leading components; we therefore fit the algorithm on the log1p-transformed, mean-centred 
data to mitigate magnitude differences \citep{box1964analysis}. The decomposition is closed-form, yielding a full set of 386 components 
that can be truncated to any target rank $K$.

NMF is a parts-based decomposition, approximating the input matrix as the product of two non-negative matrices (sectors $\times K$ and 
$K \times$ POI types), by minimising the Frobenius reconstruction error. Because each POI type's cross-sector variance is approximately 
equal to its mean, we applied a square-root transformation to stabilise variance such that high- and low-count types have similar 
variance magnitudes and contribute comparably \citep{anscombe1948transformation}. Unlike PCA, the factorisation changes depending on the choice 
of K, and must be refit for each target rank.

We tested the predictive power of the resulting $K$-dimensional embeddings for $K=5,10,15,20,50$, spanning low- to mid-rank 
embeddings, and compared the results to models that use the raw 386-dimensional features. Table~\ref{tab4} summarises the two methods and 
the assumptions they encode.

\begin{table}[!htbp]
\TBL{\caption{Comparison of the two feature-decomposition methods and the assumptions they encode.\label{tab4}}}
{\small\renewcommand{\arraystretch}{1.25}%
\begin{center}
\begin{tabular}{@{}L{1.4cm}C{2.6cm}L{3.0cm}L{2.1cm}L{2.2cm}@{}}\toprule
\TCH{Method} & \TCH{Input transformation} & \TCH{Structural constraint} & \TCH{Loading sign} & \TCH{$K$ dependence} \\\midrule
PCA & $\log(1+\text{counts})$ & Linear, orthogonal & $\pm$ & Truncate full 386 \\
NMF & $\sqrt{\text{counts}}$ & Non-negative, additive & $+$ only & Refit per $K$ \\\botrule
\end{tabular}
\end{center}}
\end{table}

\subsubsection{Modelling and Evaluation}

Using the PyCaret \citep{ali2020pycaret} library in Python, we ran a sweep of regression models for each combination of decomposition method (PCA, 
NMF) and the raw POI counts, using the $K$-dimensional embeddings introduced above. We chose seven representative models spanning linear, 
ensemble, and gradient-boosting families: ridge regression (RR), Bayesian ridge regression (BRR), random forests (RF), extra trees 
(ETR), gradient boosting (GBR), light gradient boosting (LightGBM), and extreme gradient boosting (XGBoost). This yielded 77 sweep 
combinations: 11 feature configurations (the raw 386-dimensional counts plus PCA and NMF at $K=5,10,15,20,50$) $\times$ 7 models. 

Within each combination, we fit a StandardScaler on the training set and applied it to both the validation and testing sets. 
Standardisation provides stability for models that are sensitive to feature scale (such as the regularised linear models) while leaving 
the other models unaffected \citep{hastie2009elements}. We scored each combination by its $R^2$ value on the validation set:
\begin{equation}
    R^2 = 1 - \frac{\sum_{i=1}^{n} (y_i - \hat{y}_i)^2}{\sum_{i=1}^{n} (y_i - \bar{y})^2}
\end{equation}

where $y_i$ is the log income for sector i and $\hat{y}_i$ is the corresponding model prediction, $n$ is the number of sectors in the set, 
and $\bar{y}$ is the set's average log income.

We carried forward five finalists: the best performing model on the raw counts, and for each decomposition the best across all 
$K$ and the best at $K=5$ as an interpretable reference. Each finalist was hyperparameter-tuned for 100 iterations, optimising for 
$R^2$ value, on the same train/validation split. We then refit each finalist on the full training pool ($n=12,374$) and assessed 
generalisation performance.

During the testing phase we computed $R^2$ and further introduced mean absolute error (MAE) in raw Brazilian reais and on log scale, 
and absolute relative error (ARE$_{R\$}$) in raw reais:
\begin{eqnarray}
    \text{MAE}_{\log} &=& \frac{1}{n} \sum_{i=1}^{n} |y_i - \hat{y}_i| \\
    \text{MAE}_{R\$} &=& \frac{1}{n} \sum_{i=1}^{n} |e^{y_i} - e^{\hat{y}_i}| \\
    \text{ARE}_{R\$} &=& \frac{|e^{y_i} - e^{\hat{y}_i}|}{e^{y_i}}
\end{eqnarray}

The reais-scale MAE$_{R\$}$ and ARE$_{R\$}$ both take the exponential of observed and predicted log income separately, rather than 
back-transforming an average log-scale error. ARE$_{R\$}$ is a per-sector quantity, reported as a distribution across the held-out test set; 
its right tail is heavy by construction, since low-income sectors translate even small absolute errors into large relative errors.

To quantify the uncertainty of the held-out $R^2$ estimates, we applied a marginal bootstrap procedure. We subsampled the held-out test set with
replacement for $n_{\text{boot}} = 2000$ draws and for each draw, rescored the fitted model on the subsampled held-out set. The resulting distribution of $R^2$
values across the draws provided a non-parametric estimate of the sampling distribution, from which we constructed 95\% confidence intervals
for the held-out $R^2$ of each finalist.

Finally, we assessed the spatial autocorrelation of the held-out residuals using Moran's I statistic \citep{moran1950notes}, which quantifies the degree 
of spatial clustering in the residuals. We used row-standardised 8-nearest neighbour weights based on the centroid locations of the held-out sectors, 
and reported the Moran's I value along with its associated two-sided p-value to determine the statistical significance under 999 permutations of the 
residual labels. A significant positive Moran's I indicates more clustering than expected under spatial randomness, while a significant negative 
value indicates more dispersion.

The analysis pipeline described above was implemented in Python by the authors with coding assistance from Claude Opus 4.6, 4.7, and 4.8
(Anthropic), accessed through the Claude desktop application between February and June 2026 (see Acknowledgements). The models were prompted
to help draft, debug, and refine the analysis scripts; they were not used to select the modelling approach, the validation design, or the
reported results, and every script was reviewed and executed by the authors. The full pipeline is available in the repository given in the
Data Availability Statement, so all reported figures can be reproduced independently of the tools used to write the code.

\section{Results}

Below we report the top-performing PyCaret model sweep on the held-out test set, showing a consistent spread of performance across 
decomposition methods and $K$ values, followed by validation results and an analysis of held-out residuals for the top-performing model. 
Finally, we present the top decomposition components at $K=5$, demonstrating the interpretability of the resulting embeddings and their 
activation across the municipality.

\subsection{Held-out Performance}

\begin{figure}[htbp]
\centering
\includegraphics[width=\linewidth]{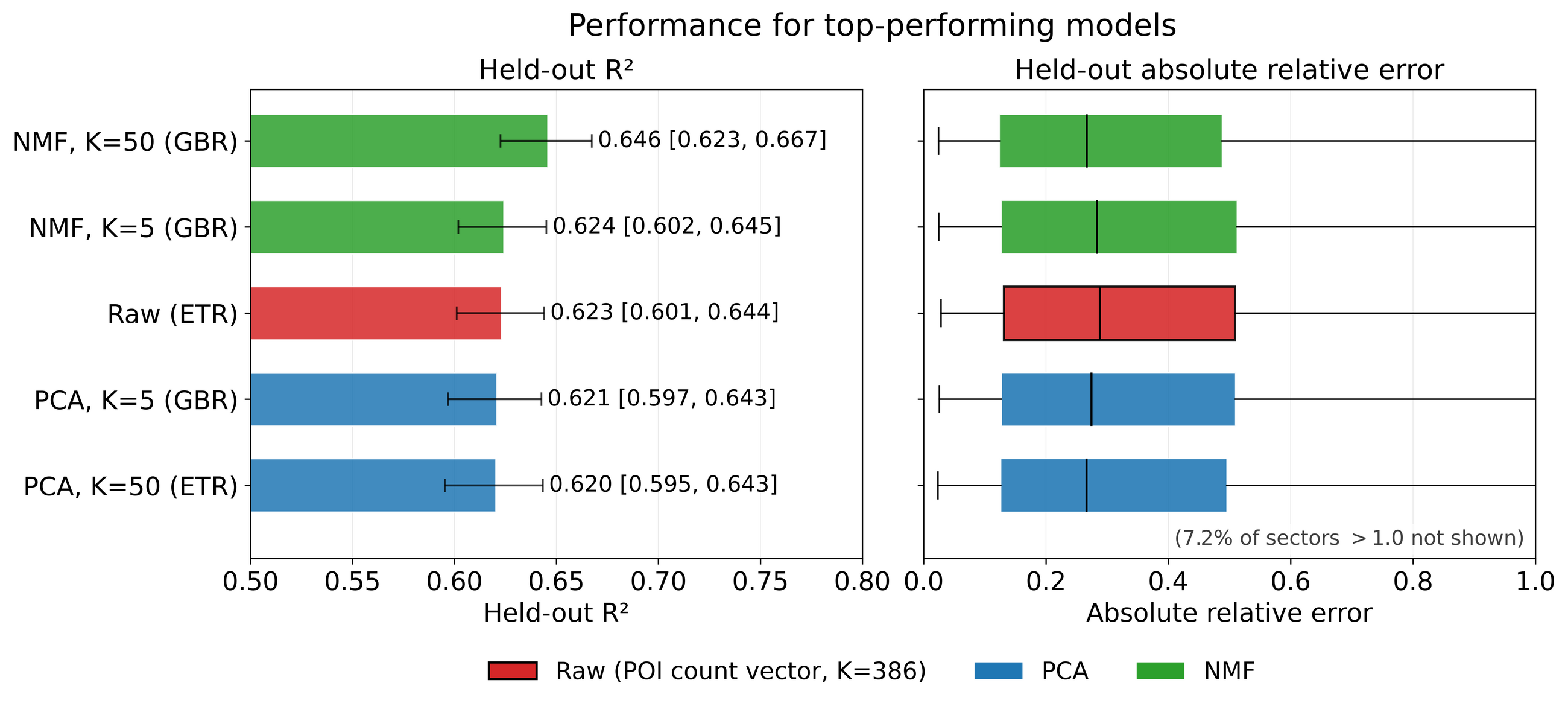}
\caption{Performance on the held-out set for the five finalists, selected from the validation stage. The left panel shows $R^2$ with the corresponding 95\% confidence interval range around the mean, computed using the bootstrap method. The right panel shows ARE$_{R\$}$, and the box plots show the distribution of values across all held-out sectors. Models are coloured according to their decomposition method, and ordered based on average $R^2$ value\label{fig4}}
\end{figure}

Figure~\ref{fig4} reports the $R^2$ and ARE$_{R\$}$ scores on the held-out test sectors for the five best-performing models selected during 
validation. Results are ordered by their mean $R^2$ value; the two NMF finalists (both with GBR) rank highest, and the top configuration 
(NMF, $K=50$, GBR) achieves a held-out $R^2$ of 0.646 (95\% CI $[0.62, 0.67]$). The bootstrap confidence intervals for mean $R^2$ all 
overlap with a spread of 0.026 across the averages. ARE$_{R\$}$ shows a similar pattern across boxplot quantiles. The held-out ordering 
differs from the validation leaderboard, where PCA at $K=50$ scored highest, though we note that given the overlapping confidence intervals, 
the ordering among the finalists should not be over-interpreted. Appendix A contains the complete validation metrics for a representative 
subset of the model-configuration combinations.

\subsection{Performance of the Top Finalist}

\begin{figure}[htbp]
\centering
\includegraphics[scale=0.15]{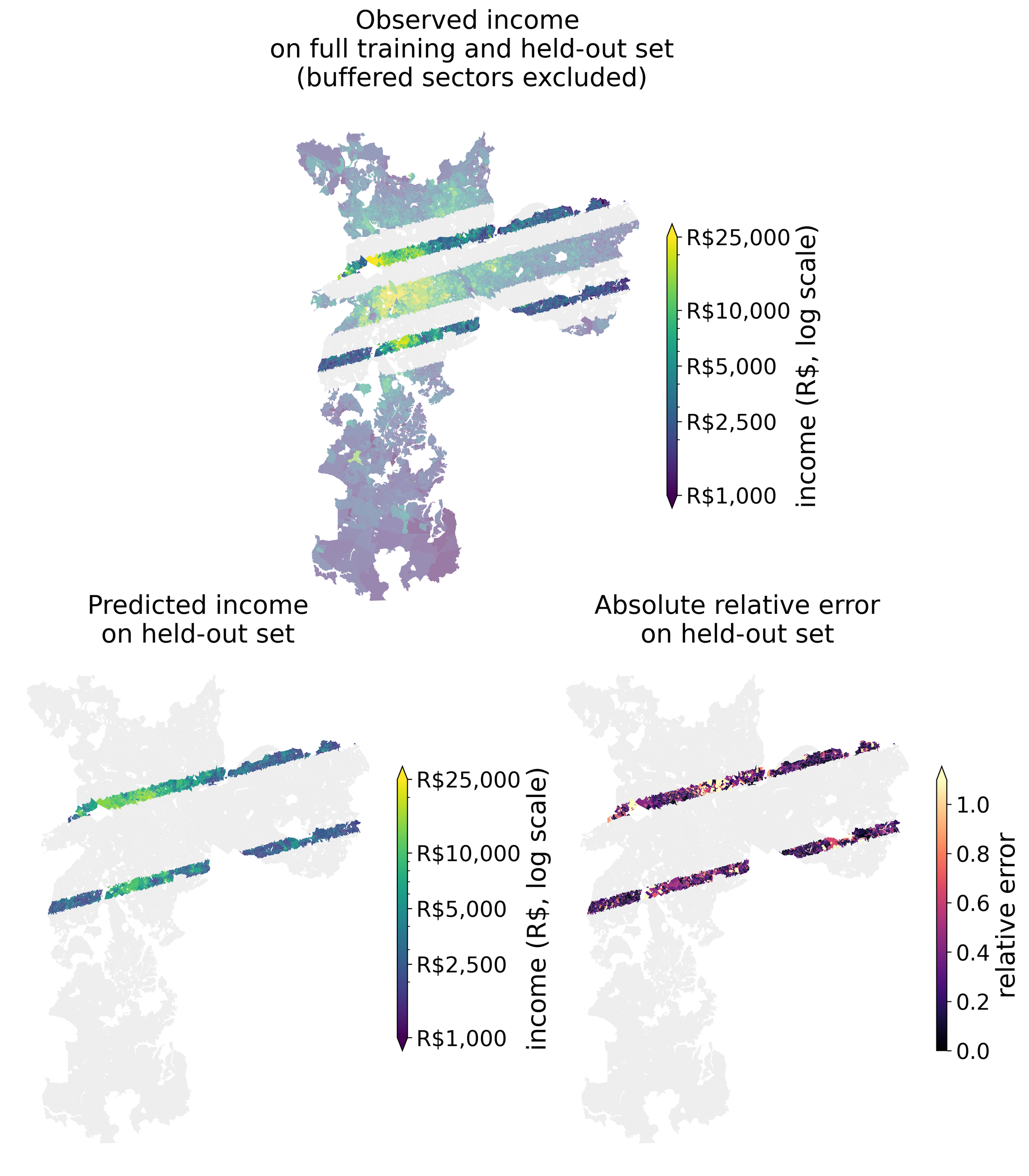}
\caption{Map of model predictions for top finalist (NMF, $K=50$, GBR). Top panel: observed income on the training set (transparent sectors) and held-out test set (buffer excluded), on a log colour scale. Bottom left panel: predicted income on the held-out set; and bottom right panel: absolute relative error of the predictions\label{fig5}}
\end{figure}

\begin{figure}[htbp]
\centering
\includegraphics[scale=0.15]{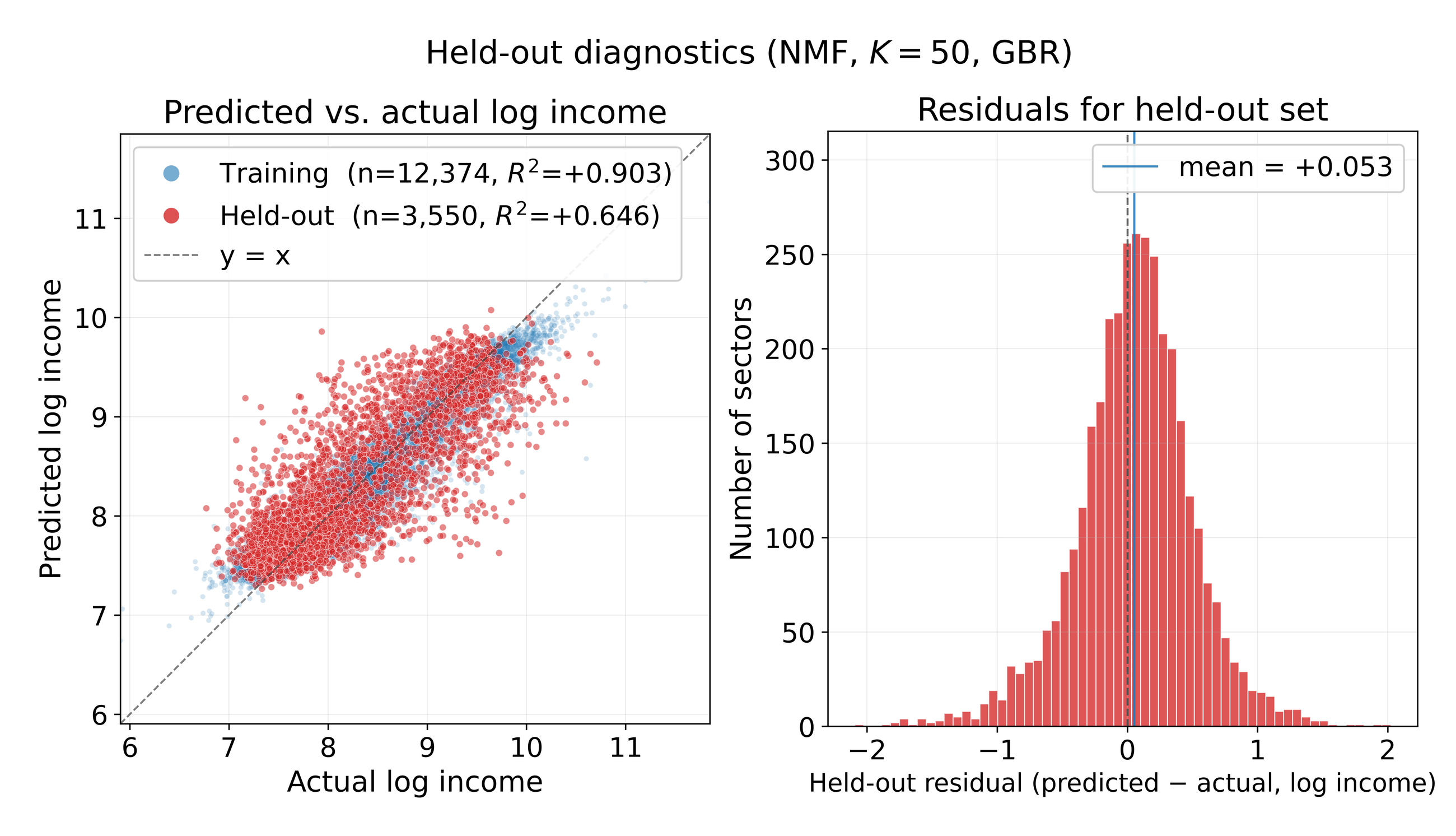}
\caption{Diagnostics on the held-out sectors (NMF, $K=50$, GBR). The left panel shows predicted vs.\ observed log income, with the held-out set in red and training set in blue. The right panel shows the distribution of the held-out residuals\label{fig6}}
\end{figure}

Figure~\ref{fig5} shows the income predictions using the NMF ($K=50$), GBR model on the held-out test set (lower left panel), along with the observed 
income (top panel) and the absolute relative error (lower right panel). The model captures the broad spatial distribution of income across 
the city, but under-predicts the highest income sectors and over-predicts the lowest income sectors, as shown in the predicted-vs-observed 
scatter plot (Figure~\ref{fig6} left). The residuals (Figure~\ref{fig6} right) are roughly symmetric around zero but heavier-tailed than a normal distribution, 
indicating that the model has more extreme errors than would be expected under a normal error assumption. Consistent with this, the model 
performs best in the mid-income range and visually compresses predictions toward the middle (Figure~\ref{fig5}).

For the top finalist, the held-out residuals have a global Moran's $I=0.334$ ($z=42.4$, two-sided $p = 0.001$). The residuals are thus more 
spatially clustered than would be expected under spatial randomness, suggesting that POI composition alone misses relevant spatial information 
and leaves systematic, spatially structured prediction error.

\subsubsection{Interpreting the Components at $K=5$}

\begin{figure}[htbp]
\centering
\includegraphics[width=\linewidth]{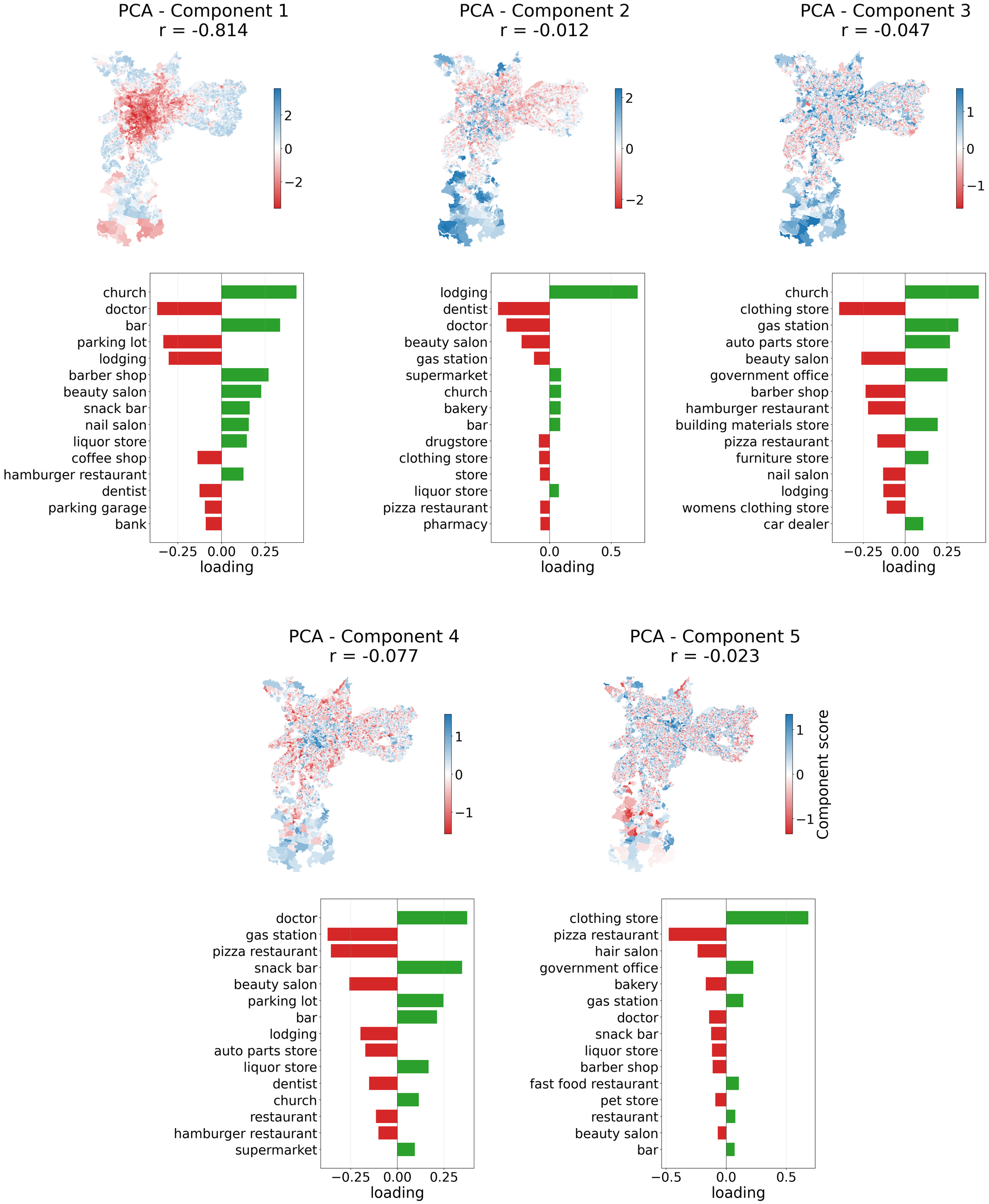}
\caption{PCA decomposition of the POI count matrix at $K=5$, fit on the training pool. The top sub-row is the spatial distribution of each component score across the municipality; the bottom sub-row shows the top POI types that load on each component\label{fig7}}
\end{figure}

\begin{figure}[htbp]
\centering
\includegraphics[width=\linewidth]{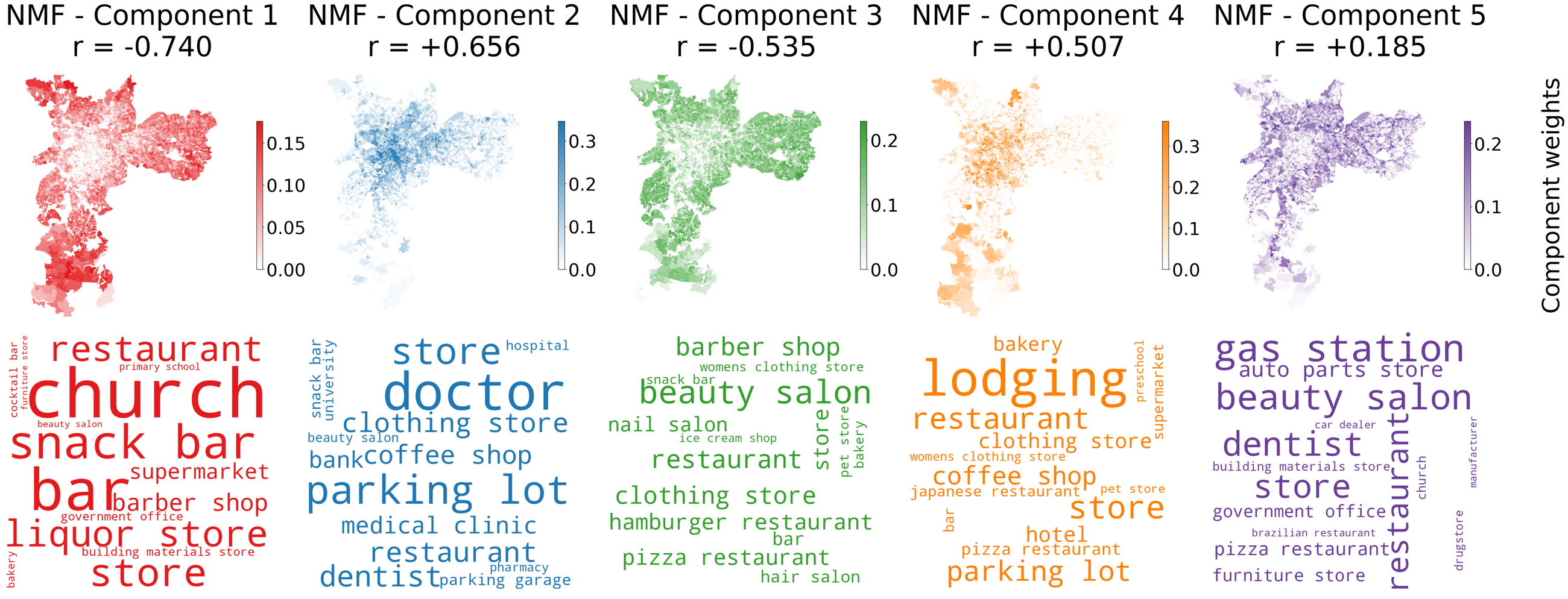}
\caption{NMF decomposition of the POI count matrix at $K=5$, fit on the training pool. The top row is the spatial distribution of each component score across the municipality; the bottom sub-row shows the top POI types that contribute to each component, with the size of each POI indicating the magnitude of contribution\label{fig8}}
\end{figure}

Choosing $K=5$ yields compact embeddings that are easier to interpret. Figure~\ref{fig7} and Figure~\ref{fig8} illustrate the PCA and NMF decomposition results 
at $K=5$. Note that this is not a per-method comparison, but an examination of the correlation between log income and the top components at a 
fixed $K$. In each figure, the top row maps show the spatial distribution of component scores across the city, indicating how strongly each 
component is activated in different sectors. The bar charts for PCA and the word clouds for NMF show the top 15 POI types loading on each 
component. The magnitude of the POI type loadings is indicated by the size of the label in the word clouds for NMF and by the absolute loading 
magnitude in the bar plots for PCA.

Since PCA loadings can be positive or negative, a sector's score reflects which sign of loadings it activates. For PCA, the first component 
has the strongest correlation with log income, with correlation values less than 0.1 for the remaining components. For NMF, the non-negative 
loadings mean each component's spatial score directly reflects its activation in different sectors, and the top components capture distinct spatial 
patterns that are strongly correlated with log income.

A full analysis is beyond the scope of this paper, but two patterns stand out across both methods. First, the component most strongly correlated 
with log income (the first for both methods) loads heavily on Church and Bar POI types, which are more prevalent in lower-income sectors. 
Second, healthcare-related POI types (e.g., Doctor and Dentist) and parking lots concentrate in the more affluent areas of the city. In PCA 
these two patterns appear as opposite-sign loadings within the first component, while in NMF they instead surface as separate components. That 
both findings largely agree with known patterns of affluence and deprivation in the city lends the components some validity, and suggests the 
models are capturing meaningful signals in the POI data.

\section{Discussion}

In this work, we demonstrate that POI composition of census sector neighbourhoods predicts sub-municipal household reference person income with 
skill levels of $R^2\approx0.65$ on a spatially held-out test set (Figure~\ref{fig4}). The top models are consistent across feature extraction methods and 
model types, suggesting a stable signal; the bootstrapped 95\% confidence intervals of the mean $R^2$ further support this, with the intervals 
overlapping for the top models and interval widths less than 0.05. 

Notably, the NMF and PCA embeddings show similar performance to the full, raw count vectors, and even compact $K=5$ embeddings retain most of 
the predictive signal. These results indicate that a low-dimensional latent signal carries much of the income-related information. Figures~\ref{fig7} and~\ref{fig8} show common patterns across the decomposition methods, including a negative correlation between income and Church, and a positive 
correlation between income and health-related POIs. The former is consistent with evidence that religious organisations cluster in lower income 
areas \citep{cheng2023spatial}. A well-documented Brazilian literature reinforces this, tracking the territorial expansion of evangelical Christianity 
into peripheral, lower-income areas: using administrative establishment data, \citet{denegri2023crescimento} document a sharp rise 
in evangelical establishments over two decades; \citet{araujo2023surgimento} traces their spread using georeferenced building records; and \citet{alves2017distribuicao} show the evangelical-to-Catholic shift is spatially clustered in peripheral districts. The latter aligns with findings that high-complexity 
healthcare facilities are more spatially accessible to high-income individuals in Brazilian cities than low-income individuals 
\citep{tomasiello2024racial}, as a high density of health-related POIs may contain a broader range of facility types. Parking-related POIs, which 
also load with affluence, are harder to read as a household-income proxy: they more plausibly mark destinations, such as clinics, service hubs, 
commercial areas, that draw car-owning visitors from elsewhere, so a sector's parking density may reflect what it offers to outsiders as much 
as who lives there. These patterns thus support the validity of the latent features and a POI-based prediction model.

Analysis of the top-performing model reveals systematic errors, including over- and under-prediction at the lower and upper tails of the income 
distribution, and statistically significant, spatially correlated errors (Figures~\ref{fig5} and~\ref{fig6}). These results suggest that POIs are missing spatial 
patterns relevant for income prediction and that the POI proxy is least effective in areas of very high and very low affluence. The inconsistency 
of POI data likely contributes to these findings. In affluent, high-density areas, the 60-POI-per-sector cap may be insufficient to fully 
characterise the commercial and service makeup of the neighbourhood. In contrast, in the lower-affluence, low-density areas, POIs are sparse, resulting 
in low-count, noisy input signals. We flag this tail-level error pattern as a direction for future work rather than a settled finding. A systematic 
comparison of where POI-based and image-based income proxies err across the income distribution, and potentially a combined approach that draws 
on each method's comparative strengths, is a natural next step, since the two data sources plausibly capture different signals at the extremes 
and in the middle. 

More broadly, our POI-based approach parallels a move in the Economic Complexity literature. That research infers a territory's social 
development, including its institutional quality and income inequality, from the structure of what it produces: how diverse and specialised 
its economic output is \citep{hartmann2017linking,hidalgo2023policy,penalvasaia2025economic}. This paper makes a structurally similar inference from 
a different kind of structure: the mix of institutions and organisations physically present in a place, such as schools, clinics, churches, 
and shops. The two are not the same object, and we do not treat them as interchangeable, but both read social and economic conditions off what 
is observably present rather than measuring income directly. Economic complexity has already been extended well below the national level to 
Brazilian states and microregions \citep{herrera2021economic,teixeira2022economic} so regional and municipal complexity 
is an established extension rather than a novel one; what has not been attempted is application at the sub-municipal, intra-city scale this 
paper occupies.

\subsection{Limitations and Future Work}

We note several significant limitations. We currently restrict analysis to income, which is an imperfect welfare measure. \citet[pp.~70--71]{sen1999development} identified the conversion factor problem: identical income levels produce different well-being levels depending on local infrastructure and 
context. \citet{alkire2011counting} operationalised this insight into a multidimensional measurement framework. We use income because it is 
the most readily available and well-validated benchmark variable for the S\~{a}o Paulo municipal context. Future work includes assessing the 
predictive power of POI-based embeddings for other welfare measures. The extension to dimensions such as education, health, and living 
standards --- and ultimately to a capabilities-grounded multidimensional deprivation index --- is the natural next phase of this research program. 

A further limitation concerns the temporal relationship between the POI data and the income benchmark. The Google Places data was collected in 
2026, while the census income data derives from Brazil's 2022 Censo Demogr\'{a}fico. The 2022 census itself was not a single moment of observation: 
fieldwork began in August 2022 and proceeded municipality by municipality over several months. The POI data introduces its own temporal imprecision: 
crowd-sourced entries are updated asynchronously, so the 2026 dataset is an accumulation of contributions made over months or years, 
potentially reflecting businesses that have since closed, relocated, or changed category. Neither variable is thus anchored to a clean 
temporal reference point. We acknowledge this compound displacement and argue that it is not a fatal flaw but a condition of working with any 
real-world combination of administrative and crowd-sourced data at scale. 

The Google Places API was attractive relative to open-source platforms such as OSM because of its better coverage. However, that coverage 
reflects businesses' incentive to be discoverable rather than the needs of the surrounding neighbourhoods, which may contribute to the 
inconsistency of POI data across income and density. Addressing these gaps is a promising avenue for future work, through supplemental signals 
such as transportation data or satellite imagery and through spatial models. 

Finally, our analysis is restricted to the S\~{a}o Paulo municipality. While S\~{a}o Paulo's income variation makes it a demanding test case, future 
work includes extending the analysis to other contexts. 

\subsection{Conclusion}

The POI input signal offers an attractive basis for predicting income: a single data source, standardised API access, and simple tabular 
outputs. Combined with the low computational costs, we present it as a reproducible option readily adoptable by under-resourced policy 
offices to inform resource allocation and programme evaluation. Unlike census-driven poverty maps fixed at enumeration, a POI-based proxy 
can be re-queried during intercensal periods for low-cost, updated estimates. However, before adoption, this approach should be robustly 
compared to higher-dimensional imagery-based black-box models. The interpretability of the present method may justify adopting it if the
performance trade-off is absent or negligible.

\section*{Appendix A}
\label{app:supplementary}

In Table~\ref{tab5} we report the validation metrics for the top ten and bottom ten configurations of the 77 evaluated; configurations are sorted by 
validation $R^2$. Table~\ref{tab6} displays the metrics on the held-out test set for the top five finalists (best performing model on the raw counts, 
and for each decomposition the best across all $K$ and the best at $K=5$ based on validation $R^2$). We introduce the mean absolute relative error 
MARE$_{R\$}$ metric, defined as the average of the per-sector ARE$_{R\$}$ values, to further contextualise the raw MAE (R\$):
\begin{equation}
    \text{MARE}_{R\$} = \frac{1}{n} \sum_{i=1}^{n} \frac{|e^{y_i} - e^{\hat{y}_i}|}{e^{y_i}}
\end{equation}

where $y_i$ is the log income for sector $i$, $\hat{y}_i$ is the corresponding model prediction, and $n$ is the number of sectors in the set.

PCA-based configurations dominate 8 of the top 10 based on validation $R^2$, but show signs of overfitting since they rank in the bottom 
two on the held-out test set.

\begin{table}[!htbp]
\TBL{\caption{Validation metrics for the top ten and bottom ten of the 77 evaluated configurations, sorted by validation $R^2$.\label{tab5}}}
{\small\setlength{\tabcolsep}{5pt}\renewcommand{\arraystretch}{1.2}%
\begin{center}
{%
\begin{tabular}{@{}lllrrrr@{}}\toprule
\TCH{Method} & \TCH{$K$} & \TCH{Estimator} & \TCH{Val.\ $R^2$} & \TCH{Val.\ RMSE} & \TCH{Val.\ MAE (log)} & \TCH{Val.\ MAE (R\$)} \\\midrule
PCA & 50 & ETR & 0.712 & 0.397 & 0.301 & 1,531 \\
PCA & 50 & LightGBM & 0.704 & 0.402 & 0.304 & 1,564 \\
PCA & 50 & GBR & 0.704 & 0.402 & 0.306 & 1,568 \\
Raw & 386 & ETR & 0.704 & 0.402 & 0.307 & 1,591 \\
PCA & 20 & GBR & 0.702 & 0.404 & 0.306 & 1,578 \\
PCA & 50 & BRR & 0.701 & 0.404 & 0.310 & 1,614 \\
PCA & 50 & RR & 0.701 & 0.404 & 0.310 & 1,615 \\
PCA & 15 & GBR & 0.701 & 0.404 & 0.307 & 1,583 \\
NMF & 50 & GBR & 0.701 & 0.404 & 0.305 & 1,575 \\
PCA & 20 & ETR & 0.700 & 0.405 & 0.308 & 1,586 \\\midrule
\multicolumn{7}{c}{57 mid-ranked configurations omitted} \\\midrule
NMF & 5 & RF & 0.663 & 0.429 & 0.325 & 1,685 \\
PCA & 10 & XGBoost & 0.663 & 0.429 & 0.330 & 1,742 \\
NMF & 5 & ETR & 0.662 & 0.430 & 0.328 & 1,704 \\
PCA & 5 & XGBoost & 0.661 & 0.431 & 0.331 & 1,770 \\
PCA & 15 & XGBoost & 0.653 & 0.435 & 0.333 & 1,732 \\
NMF & 50 & XGBoost & 0.649 & 0.438 & 0.330 & 1,730 \\
NMF & 5 & XGBoost & 0.648 & 0.439 & 0.331 & 1,729 \\
NMF & 20 & XGBoost & 0.648 & 0.439 & 0.334 & 1,758 \\
NMF & 10 & XGBoost & 0.645 & 0.441 & 0.336 & 1,741 \\
NMF & 15 & XGBoost & 0.635 & 0.447 & 0.343 & 1,787 \\\botrule
\end{tabular}}
\end{center}}
\end{table}

\begin{table}[!htbp]
\TBL{\caption{Held-out test-set metrics for the top five finalists.\label{tab6}}}
{\small\setlength{\tabcolsep}{3.5pt}\renewcommand{\arraystretch}{1.2}%
\begin{center}
{%
\begin{tabular}{@{}lllrrrrr@{}}\toprule
\TCH{Method} & \TCH{$K$} & \TCH{Estimator} & \TCH{Test $R^2$} & \TCH{Test RMSE} & \TCH{Test MAE (log)} & \TCH{Test MAE (R\$)} & \TCH{Test MARE (R\$)} \\\midrule
NMF & 50 & GBR & 0.646 & 0.455 & 0.343 & 1,928 & 0.386 \\
NMF & 5 & GBR & 0.624 & 0.468 & 0.357 & 1,963 & 0.406 \\
Raw & 386 & ETR & 0.623 & 0.469 & 0.358 & 1,959 & 0.397 \\
PCA & 5 & GBR & 0.621 & 0.471 & 0.356 & 1,994 & 0.413 \\
PCA & 50 & ETR & 0.620 & 0.471 & 0.353 & 1,985 & 0.404 \\\botrule
\end{tabular}}
\end{center}}
\end{table}

\newpage

\section*{Declarations}
\addcontentsline{toc}{section}{Declarations}

\subsection*{Acknowledgements}
The authors made use of Claude Opus 4.6, Claude Opus 4.7, and Claude Opus 4.8 (Anthropic), accessed through the Claude desktop application
(\url{https://claude.ai}) between February and June 2026, to assist with planning and refining the structure of the manuscript and with
drafting, debugging, and improving the analysis code. The models were prompted to suggest an outline and organisation for the manuscript's
sections and to help draft, debug, and refine the analysis scripts described in Section~\ref{sec:methods}. These tools were not used to generate 
figures or to produce any of the reported results. The original conception of the
study and the analysis plan were developed by the authors, who reviewed, edited, and verified all resulting text and code. The authors are
entirely responsible for the scientific content of this paper and confirm that it adheres to the journal's authorship policy.

\subsection*{Funding Statement}
This work was supported by a seed grant from the Hastings Initiative for AI and Humanity at Bowdoin College. The funding source had no role in the study design; 
in the collection, analysis, or interpretation of the data; or in the writing of the article and the decision to submit it for publication.

\subsection*{Competing Interests}
The authors declare no conflicting interests with respect to the research, authorship, or publication of this article.

\subsection*{Data Availability Statement}
The code and data supporting the findings of this study are openly available in a GitHub
repository at \href{https://github.com/Geolocated-Poverty-Prediction/saopaulo-poi-income-prediction}{https://github.com/Geolocated-Poverty-Prediction/saopaulo-poi-income-prediction}.

\subsection*{Author Contributions}
\noindent Adrienne C. Kinney: Conceptualization, Data curation, Formal analysis, Investigation, Methodology, Validation, Visualization, Writing -- original draft, Writing -- review \& editing.

\noindent Anya Workman: Conceptualization, Data curation, Investigation, Validation, Writing -- original draft, Writing -- review \& editing.

\noindent Ademar Takeo Akabane: Writing -- review \& editing.

\noindent Jenna Barac: Data curation, Investigation, Visualization, Writing -- review \& editing.

\noindent Paulo Fernando Braga Carvalho: Writing -- review \& editing.

\noindent Jeova Farias: Conceptualization, Formal analysis, Methodology, Project administration, Supervision, Validation, Writing -- review \& editing.

\noindent Fernando Nascimento: Conceptualization, Formal analysis, Methodology, Project administration, Supervision, Validation, Writing -- original draft, Writing -- review \& editing.

\noindent Paulo Ricardo da Silva Oliveira: Writing -- review \& editing.

\bibliographystyle{apalike}
\bibliography{POI_SaoPaulo}

\end{document}